\documentclass[a4paper,11pt]{article}

\usepackage{hyperref}%
\hypersetup{colorlinks=true,%
  linkcolor=blue,%
  citecolor=magenta,%
  urlcolor=black,%
  bookmarksnumbered=true,%
  bookmarkstype=toc,%
  bookmarksopen=false,%
  pdftitle={Emergent Anyon Distribution in the Unruh Effect},%
  pdfkeywords={Unruh effect; anyon; conformal field theory},%
  pdfauthor={Satoshi Ohya}}

\usepackage[margin=1in]{geometry}

\usepackage[bitstream-charter,cal=cmcal]{mathdesign}
\usepackage[T1]{fontenc}

\usepackage{xcolor}

\usepackage{amsmath,amsbsy,bm}

\usepackage{epic,eepic}

\usepackage[hang,font=small,labelfont=bf]{caption}

\usepackage{titlesec}
\titleformat{\section}[block]{\filright\bfseries}{\thesection.}{0.5em}{}[]
\titleformat{\subsection}[block]{\filright\bfseries}{\thesubsection.}{0.5em}{}

\usepackage{cite}

\title{\large\bfseries Emergent Anyon Distribution in the Unruh
  Effect}
\author{
  {\normalsize Satoshi Ohya}\\[1em]
  {\small\textit{Institute of Quantum Science, Nihon University}}\\
  {\small\textit{Kanda-Surugadai 1-8-14, Chiyoda, Tokyo 101-8308,
      Japan}}\\[1ex]
  {\small\texttt{ohya@phys.cst.nihon-u.ac.jp}}}
\date{\small (Dated: \today)}

\begin{document}
\maketitle
\flushbottom

\begin{abstract}
  We point out that, when the Unruh--DeWitt detector couples to a
  scalar primary operator of $d$-dimensional conformal field theory,
  the detector's power spectrum generally obeys the thermal
  distribution for $(1+1)$-dimensional anyons.
\end{abstract}

\section{Introduction}
\label{section:1}
A uniformly accelerating observer of constant proper acceleration
$a(>0)$ sees the vacuum for an inertial observer as a thermal state of
temperature $T=a/(2\pi)$. This is the Unruh effect \cite{Unruh:1976db}
and has been vastly studied over the last forty years both from
physical and mathematical perspectives. From the physical viewpoint,
the Unruh effect provides a theoretical laboratory for understanding
quantum phenomena in gravitational field background, whose typical
example is the Hawking radiation of black hole
\cite{Hawking:1974sw}. It also provides a nice pedagogical example of
Takahashi--Umezawa thermofield dynamics \cite{Takahashi:1996zn}---an
interplay between entanglement and thermalization. From the
mathematical viewpoint, on the other hand, the Unruh effect provides a
nice geometrical realization of Tomita--Takesaki modular theory of
operator algebras, in which the modular operator and modular
conjugation are simply given by the complexified Lorentz boost and CPT
conjugate (with partial reflection), respectively. It also provides
the interplay between the Bisognano--Wichmann theorem
\cite{Bisognano:1975ih,Bisognano:1976za} and the
Kubo--Martin--Schwinger (KMS) thermal equilibrium condition
\cite{Haag:1967sg}, based on which the Unruh effect can be proved
mathematically. Though it is shown axiomatically by Sewell
\cite{Sewell:1982zz} that correlation functions with respect to the
inertial vacuum satisfy the KMS conditions with respect to the Lorentz
boost, whereby the thermality of inertial vacuum for accelerating
observers is rigorously proved, there still remain puzzles in the
Unruh effect. Among them is the Takagi's \textit{statistics inversion}
\cite{Takagi:1986kn}, which is an apparent phenomenon that
interchanges the Bose--Einstein and Fermi--Dirac distributions. In the
mid-1980s Takagi studied the power spectrum of the Unruh--DeWitt
detector \cite{DeWitt:1979} coupled to a massless scalar field in
generic spacetime dimension $d$ and found that, when $d$ is even, the
detector's power spectrum obeys the Bose--Einstein distribution. This
is as naively expected since the power spectrum is simply given by the
Fourier transform of scalar two-point Wightman function. However, when
$d$ is odd, he found that the power spectrum obeys the Fermi--Dirac
distribution even without any fermionic degrees of freedom. Soon after
this discovery, Ooguri pointed out \cite{Ooguri:1985nv} that this
puzzle can be explained in terms of the absence of Huygens principle
in odd dimensional spacetimes \cite{Courant:1962}. In other words,
this apparent phenomena can be understood from the analytic structure
of two-point Wightman functions and resulting anti-commutativity for
time-like separated scalar correlators, which can happen only for
massless non-interacting fields in odd dimensions. Instead of massless
free fields, in this short note we shall consider a primary operator
of conformal field theory, for which the Huygens principle is violated
as well. The goal of this note is to show that, when the Unruh--DeWitt
detector couples to a scalar primary operator of conformal field
theory with non-vanishing anomalous dimension, the detector's power
spectrum generally obeys the thermal distribution for
$(1+1)$-dimensional anyons that is derived by Liguori, Mintchev, and
Pilo in 1999 by making use of the bosonization technique
\cite{Liguori:1999tw}. We shall see that the Takagi's statistics
inversion is then generalized to a \textit{statistics
  interpolation}---an apparent phenomenon which interpolates the
Bose--Einstein and Fermi--Dirac distributions in a continuous
way.\footnote{\label{footnote:1}Yet another statistics interpolation
  was discussed by Sriramkumar \cite{Sriramkumar:2002dn}.} The scaling
dimension of primary operator turns out to play the role of the
so-called statistical parameter. Before embarking on this statistics
interpolation, however, let us first briefly recall the Takagi's
findings.

\section{Statistics Inversion: A Review}
\label{section:2}
The central objects of statistics inversion are the so-called
Unruh--DeWitt detector introduced by DeWitt \cite{DeWitt:1979} and its
power spectrum (also known as the response function). Let us consider
a uniformly accelerating point-like detector in the $d$-dimensional
Minkowski spacetime. Let us further suppose that the whole spacetime
is filled with a quantum field theory. The combined system of detector
and quantum field theory is then schematically described by the action
\begin{align}
  S_{\text{total}}=S_{\text{detector}}+S_{\text{QFT}}+S_{\text{int}},\label{eq:1}
\end{align}
where $S_{\text{detector}}$ and $S_{\text{QFT}}$ are the actions of
detector and background quantum field theory whose explicit forms are
not necessary in the following discussions. For simplicity we shall
focus on a hermitian scalar field as the background. DeWitt considered
that the detector has a ``monopole moment'' $m(\tau)$ and interacts
with the background scalar field $\phi(x)$ via the following term:
\begin{align}
  S_{\text{int}}=\int_{-\infty}^{\infty}\!\!\!d\tau\,m(\tau)\phi(x(\tau)),\label{eq:2}
\end{align}
where $\tau$ is the detector's proper time and $x^{\mu}(\tau)$ is the
worldline for the uniformly accelerating detector given by
\begin{align}
  x^{0}(\tau)=\frac{1}{a}\sinh(a\tau),\quad
  x^{1}(\tau)=\frac{1}{a}\cosh(a\tau),\quad
  \bm{x}_{\perp}(\tau)=\text{constant}.\label{eq:3}
\end{align}
We wish to know the transition amplitude for the process
$|E_{i}\rangle\otimes|0\rangle\to|E_{f}\rangle\otimes|\Phi\rangle$,
where $|E_{i/f}\rangle\in\mathcal{H}_{\text{detector}}$ is the initial
(final) state of the detector and
$|0\rangle\in\mathcal{H}_{\text{QFT}}$ is the vacuum for inertial
observers. The final state $|\Phi\rangle\in\mathcal{H}_{\text{QFT}}$
is an arbitrary basis of background quantum field theory which is
assumed to satisfy the completeness
$\sum_{\Phi}|\Phi\rangle\langle\Phi|=1$. In the interaction picture,
the transition amplitude is given by the matrix element
$A(E_{f},\Phi|E_{i},0)=\langle
E_{f},\Phi|(T\exp(iS_{\text{int}})-1)|E_{i},0\rangle$ which, at the
linear order of perturbation theory, takes the following form:
\begin{align}
  A(E_{f},\Phi|E_{i},0)
  \simeq i\langle E_{f}|m(0)|E_{i}\rangle
  \int_{-\infty}^{\infty}\!\!\!d\tau\,\mathrm{e}^{i(E_{f}-E_{i})\tau}\langle\Phi|\phi(x(\tau))|0\rangle.\label{eq:4}
\end{align}
The transition probability per unit time---the transition rate---is
then given by summing over all possible final states:
\begin{align}
  R(E_{f}|E_{i})
  =\frac{1}{T}\sum_{\Phi}\left|A(E_{f},\Phi|E_{i},0)\right|^{2}
  \simeq\left|\langle E_{f}|m(0)|E_{i}\rangle\right|^{2}F(\omega),\label{eq:5}
\end{align}
where $T=\int_{-\infty}^{\infty}\!d\tau$ is the total time for the
process. Here $F(\omega)$ is the so-called detector's power spectrum
and given by the Fourier transform of two-point Wightman function
along the detector's worldline:
\begin{align}
  F(\omega)=\int_{-\infty}^{\infty}\!\!\!d\tau\,\mathrm{e}^{-i\omega\tau}\langle0|\phi(x(\tau))\phi(x(0))|0\rangle,
  \quad
  \omega=E_{f}-E_{i}.\label{eq:6}
\end{align}
Let us now focus on a massless free scalar field $\phi(x)$. In
$d$-dimensional spacetime $\phi(x)$ has the mass dimension $(d-2)/2$
and hence its two-point Wightman function
$\langle0|\phi(x)\phi(y)|0\rangle$ is basically given by the power
function $z^{-(d-2)/2}$ of the Poincar\'{e} invariant
$z=(x-y)^{2}=-(x^{0}-y^{0})^{2}+|\bm{x}-\bm{y}|^{2}$. There is,
however, a subtlety if $d$ is odd and $\phi(x)$ and $\phi(y)$ are
time-like separated because in this case $z^{-(d-2)/2}$ has the branch
cut along the negative $z$-axis and hence its value depends on how one
approaches the negative $z$-axis. A careful analysis shows that
$z=(x-y)^{2}$ should be chosen as $z=|z|\mathrm{e}^{i\pi}$ for
$(x-y)^{2}<0$ and $x^{0}-y^{0}>0$ and as $z=|z|\mathrm{e}^{-i\pi}$ for
$(x-y)^{2}<0$ and $x^{0}-y^{0}<0$.\footnote{\label{footnote:2}This is
  of course consistent with the $i\epsilon$ prescription: the Wightman
  function \eqref{eq:7} is the boundary value (i.e., the limit
  $\epsilon\to+0$) of the complex function
  $\langle0|\phi(x)\phi(y)|0\rangle\propto[-(x^{0}-y^{0}-i\epsilon)^{2}+|\bm{x}-\bm{y}|^{2}]^{-(d-2)/2}=[(x-y)^{2}+i\epsilon\mathrm{sgn}(x^{0}-y^{0})]^{-(d-2)/2}$,
  which indicates that one has to approach the real $z$-axis from
  above (below) if $x^{0}-y^{0}>0$ ($x^{0}-y^{0}<0$). Notice that
  there is no discontinuity across the positive real $z$-axis.}
Indeed, by canonically quantizing the field $\phi(x)$ one can see that
the two-point Wightman function takes the following forms:
\begin{align}
  \langle0|\phi(x)\phi(y)|0\rangle=
  \begin{cases}
    \displaystyle
    \frac{\Gamma(\frac{d-2}{2})}{4\pi^{\frac{d}{2}}}\frac{1}{[(x-y)^{2}]^{\frac{d-2}{2}}}
    &\text{for}\quad(x-y)^{2}>0,\\[1.5em]
    \displaystyle
    \frac{\Gamma(\frac{d-2}{2})}{4\pi^{\frac{d}{2}}}\frac{\mathrm{e}^{-i\,\mathrm{sgn}(x^{0}-y^{0})\frac{d-2}{2}\pi}}{[-(x-y)^{2}]^{\frac{d-2}{2}}}
    &\text{for}\quad(x-y)^{2}<0,
  \end{cases}\label{eq:7}
\end{align}
where $\mathrm{sgn}(x)=x/|x|$ ($\mathrm{sgn}(0)=0$) is the sign
function. The power spectrum \eqref{eq:6} is evaluated by just
substituting the worldline \eqref{eq:3} into the Wightman function
\eqref{eq:7} and then Fourier transform with respect to the proper
time. To do this, one should pay attention to the following two facts:
the first is that any two points lying on the worldline \eqref{eq:3}
are always time-like separated; that is,
$(x(\tau)-x(\tau^{\prime}))^{2}=-(4/a^{2})\sinh^{2}(a(\tau-\tau^{\prime})/2)<0$
for any $\tau-\tau^{\prime}\neq0$. The second is that, since
$x^{0}(\tau)=(1/a)\sinh(a\tau)$ is a monotonically increasing function
of $\tau$, the sign function
$\mathrm{sgn}(x^{0}(\tau)-x^{0}(\tau^{\prime}))$ is equivalent to
$\mathrm{sgn}(\tau-\tau^{\prime})$. Keeping these in mind, one can
find that the Wightman function along the uniformly accelerating
detector takes the following form:
\begin{align}
  \langle0|\phi(x(\tau))\phi(x(\tau^{\prime}))|0\rangle
  =\frac{\Gamma(\frac{d-2}{2})}{4\pi^{\frac{d}{2}}}
  \left[\frac{\pi T}{i\sinh(\pi T(\tau-\tau^{\prime}))}\right]^{d-2}
  \quad\text{for}\quad\tau-\tau^{\prime}\neq0,\label{eq:8}
\end{align}
which satisfies the KMS condition
$\langle0|\phi(x(\tau))\phi(x(0))|0\rangle=\langle0|\phi(x(0))\phi(x(\tau+i/T))|0\rangle$.
As we will see shortly in the next section, the Fourier transform of
\eqref{eq:8} is exactly calculable. The result is that, when $d$ is
even, $F(\omega)$ is proportional to the Bose--Einstein distribution
and the proportional coefficient is a polynomial of $\omega$. This
signifies that the uniformly accelerating detector registers the
thermal spectrum of bosonic particles as if it were immersed in a heat
bath, which would have been a major surprise in the early days of the
field. But a real surprise is that, if $d$ is odd, $F(\omega)$ becomes
proportional to the Fermi--Dirac distribution even in the absence of
fermionic degrees of freedom. The resultant power spectrum turns out
to be of the following forms:
\begin{align}
  F(\omega)=
  \begin{cases}
    \displaystyle\frac{1}{\mathrm{e}^{\omega/T}-1}\times(\text{polynomial of $\omega$}) &\text{for $d$ even},\\[1em]
    \displaystyle\frac{1}{\mathrm{e}^{\omega/T}+1}\times(\text{polynomial of $\omega$}) &\text{for $d$ odd}.\\
  \end{cases}\label{eq:9}
\end{align}
This is the Takagi's statistics inversion and one of the biggest
bewildering results in the Unruh effect. And the reason why this
happens can be explained---at least at the mathematical level---in
terms of the Huygens principle
\cite{Takagi:1986kn,Ooguri:1985nv}.\footnote{The Huygens principle
  roughly means that solutions of massless free field equations with
  sharply localized initial states propagate only along the future
  light-cone. In quantum field theory language, this is equivalent to
  the statement that the retarded Green's function
  $i\theta(x^{0}-y^{0})\langle0|[\phi(x),\phi(y)]|0\rangle$ has a
  support only on the future light-cone
  $V_{+}=\{x^{\mu}-y^{\mu}: (x-y)^{2}=0,~x^{0}-y^{0}\geq0\}$. This is
  true only for $d$ even. For $d$ odd, the retarded Green's function
  has a support on the closed future light-cone
  $\Bar{V}_{+}=\{x^{\mu}-y^{\mu}: (x-y)^{2}\leq0,~x^{0}-y^{0}\geq0\}$;
  that is, once one hears the sound of massless free fields, the
  reverberation never disappears in odd dimensional spacetime. This is
  the absence of Huygens principle, whose origin is the
  anti-commutation relation \eqref{eq:11} for $d$ odd.} To see this,
let us first note that the power spectrum \eqref{eq:6} can be recast
into the following two equivalent forms:\footnote{To derive these, use
  the translation invariance
  $\langle0|\phi(x(0))\phi(x(\tau))|0\rangle=\langle0|\phi(x(-\tau))\phi(x(0))|0\rangle$
  and the KMS condition in frequency space,
  $F(-\omega)=\mathrm{e}^{\omega/T}F(\omega)$. For example, the
  Fourier transform of anti-commutator function is written as
  $\int_{-\infty}^{\infty}\!d\tau\,\mathrm{e}^{-i\omega\tau}\langle0|[\phi(x(\tau)),\phi(x(0))]|0\rangle=F(\omega)-F(-\omega)=(1-\mathrm{e}^{\omega/T})F(\omega)$,
  which proves eq.~\eqref{eq:10a}. Likewise, one can prove
  eq.~\eqref{eq:10b}. It should be noted that eqs.~\eqref{eq:10a} and
  \eqref{eq:10b} hold true for generic interacting fields with or
  without mass. The exchange relation \eqref{eq:11}, on the other
  hand, holds true only for massless free fields.}
\begin{subequations}
  \begin{align}
    F(\omega)
    &=-\frac{1}{\mathrm{e}^{\omega/T}-1}\int_{-\infty}^{\infty}\!\!\!d\tau\,
      \mathrm{e}^{-i\omega\tau}\langle0|\left[\phi(x(\tau)),\phi(x(0))\right]|0\rangle\label{eq:10a}\\
    &=+\frac{1}{\mathrm{e}^{\omega/T}+1}\int_{-\infty}^{\infty}\!\!\!d\tau\,
      \mathrm{e}^{-i\omega\tau}\langle0|\left\{\phi(x(\tau)),\phi(x(0))\right\}|0\rangle,\label{eq:10b}
  \end{align}
\end{subequations}
which just follow from the KMS condition. The second important point
to note is that the Wightman function \eqref{eq:7} for time-like
separated operators satisfies the following exchange relation:
\begin{align}
  \langle0|\phi(x)\phi(y)|0\rangle=\mathrm{e}^{-i\,\mathrm{sgn}(x^{0}-y^{0})(d-2)\pi}\langle0|\phi(y)\phi(x)|0\rangle
  \quad\text{for}\quad
  (x-y)^{2}<0,\label{eq:11}
\end{align}
which becomes anti-commutation relation when $d$ is odd. Hence if $d$
is odd, the anti-commutator function
$\langle0|\{\phi(x(\tau)),\phi(x(0))\}|0\rangle$ becomes zero for
$\tau\neq0$, which implies that it is non-vanishing only at
$\tau=0$. Such function with a one-point support at $\tau=0$ must be
given by a polynomial of the delta function $\delta(\tau)$ and its
derivatives, whose Fourier transform \eqref{eq:10b} must then become a
polynomial of $\omega$. Similar arguments hold true for the commutator
function $\langle0|[\phi(x(\tau)),\phi(x(0))]|0\rangle$ for $d$ even
and its Fourier transform \eqref{eq:10a}. Putting all these things
together, one arrives at the expressions \eqref{eq:9}. This is the
Ooguri's explanation for the statistics inversion
\cite{Ooguri:1985nv}.

The Huygens principle offers a nice explanation for the form of
detector's power spectrum \eqref{eq:9}, however, one may still find
this hard to understand intuitively. Yet, regardless of the presence
or absence of intuition, the computational origin of statistic
inversion is obvious: it just comes from the \textit{power} of the
two-point Wightman function \eqref{eq:7} (or \eqref{eq:8}), which is
just determined by the mass dimension of free scalar fields in $d$
dimensions. Hence it would be natural to expect that, if $\phi(x)$ had
an \textit{anomalous dimension}, the power spectrum could obey a
thermal distribution function for some generalized statistics that
interpolates the Bose--Einstein and Fermi--Dirac distributions
continuously. As we shall see from now on, this is indeed the case if
the detector's ``monopole moment'' couples to a scalar primary
operator of conformal field theory with non-vanishing anomalous
dimension.

\section{Statistics Interpolation: Emergent Anyon Distribution}
\label{section:3}
Let $\mathcal{O}_{\Delta}(x)$ be a hermitian scalar primary operator
of scaling dimension $\Delta$ and couple to the ``monopole moment''
$m(\tau)$ via the interaction term
$S_{\text{int}}=\int_{-\infty}^{\infty}\!\!d\tau\,m(\tau)\mathcal{O}_{\Delta}(x)$. At
the linear order of perturbation theory, the detector's power spectrum
is again given by \eqref{eq:6} with $\phi$ being replaced by
$\mathcal{O}_{\Delta}$. The two-point Wightman function of
$\mathcal{O}_{\Delta}(x)$ is just determined from the conformal
symmetry\footnote{For textbook exposition of $d$-dimensional conformal
  field theory, we refer to \cite{Fradkin:1996is}.} and takes the
following forms:
\begin{align}
  \langle0|\mathcal{O}_{\Delta}(x)\mathcal{O}_{\Delta}(y)|0\rangle=
  \begin{cases}
    \displaystyle
    \frac{\Gamma(2\Delta)A_{\Delta}}{2\pi}\frac{1}{[(x-y)^{2}]^{\Delta}}
    &\text{for}\quad(x-y)^{2}>0,\\[1.5em]
    \displaystyle
    \frac{\Gamma(2\Delta)A_{\Delta}}{2\pi}\frac{\mathrm{e}^{-i\,\mathrm{sgn}(x^{0}-y^{0})\pi\Delta}}{[-(x-y)^{2}]^{\Delta}}
    &\text{for}\quad(x-y)^{2}<0,
  \end{cases}\label{eq:12}
\end{align}
where $A_{\Delta}$ is a dimensionless normalization factor which is
assumed to be real. $\Gamma(2\Delta)/(2\pi)$ is factored out just for
later convenience. It should be emphasized that, as noted in footnote
\ref{footnote:2}, the appearance of phase
$\mathrm{e}^{-i\,\mathrm{sgn}(x^{0}-y^{0})\pi\Delta}$ is consistent
with the $i\epsilon$-prescription. One can easily check that
eq.~\eqref{eq:12} enjoys the following properties:
\begin{itemize}
\item \textbf{Hermiticity.} The two-point Wightman function satisfies
  the identity
  \begin{align}
    \overline{\langle0|\mathcal{O}_{\Delta}(x)\mathcal{O}_{\Delta}(y)|0\rangle}
    =\langle0|\mathcal{O}_{\Delta}(y)\mathcal{O}_{\Delta}(x)|0\rangle
    \quad\text{for}\quad
    \forall x,y\in\mathbb{R}^{1,d-1},\label{eq:13}
  \end{align}
  where the overline stands for the complex conjugate. This is just a
  manifestation of hermiticity
  $\mathcal{O}_{\Delta}^{\dagger}(x)=\mathcal{O}_{\Delta}(x)$ for the
  scalar primary operator.
  
\item \textbf{Space-like commutativity.} The two-point Wightman
  functions for space-like separated operators commute with each
  other:
  \begin{align}
    \langle0|\mathcal{O}_{\Delta}(x)\mathcal{O}_{\Delta}(y)|0\rangle
    =\langle0|\mathcal{O}_{\Delta}(y)\mathcal{O}_{\Delta}(x)|0\rangle
    \quad\text{for}\quad
    (x-y)^{2}>0.\label{eq:14}
  \end{align}
  This is just a manifestation of causality.\footnote{Strictly
    speaking, causality is an intricate problem in conformal field
    theory because \textit{finite} special conformal transformations
    generally change the sign of the Poincar\'{e} invariant
    $(x-y)^{2}$. (The maximal symmetry transformations that preserve
    the causal structure of Minkowski spacetime are only the
    Poincar\'{e} transformations and dilatation.) In order to make the
    special conformal transformations both globally well-defined and
    causal, one needs to work in the universal covering of conformally
    compactified Minkowski spacetime; see, e.g.,
    ref.~\cite{Kastrup:1975qa} for a short review. In this note we
    will not touch upon this causality issue.}
  
\item \textbf{Discontinuity across the branch cut.} The two-point
  Wightman function for time-like separated operators acquires a phase
  when $\mathcal{O}_{\Delta}(x)$ and $\mathcal{O}_{\Delta}(y)$ are
  swapped with each other:
  \begin{align}
    \langle0|\mathcal{O}_{\Delta}(x)\mathcal{O}_{\Delta}(y)|0\rangle
    =\mathrm{e}^{-i\,\mathrm{sgn}(x^{0}-y^{0})2\pi\Delta}\langle0|\mathcal{O}_{\Delta}(y)\mathcal{O}_{\Delta}(x)|0\rangle
    \quad\text{for}\quad
    (x-y)^{2}<0.\label{eq:15}
  \end{align}
  This is just a manifestation of multi-valuedness as a function of
  $z=(x-y)^{2}$ and resulting discontinuity across the branch cut
  along the negative $z$-axis.
\end{itemize}
Notice that the appearance of phase does not conflict with the
hermiticity \eqref{eq:13}.

\begin{figure}[t]
  \centering
  \input{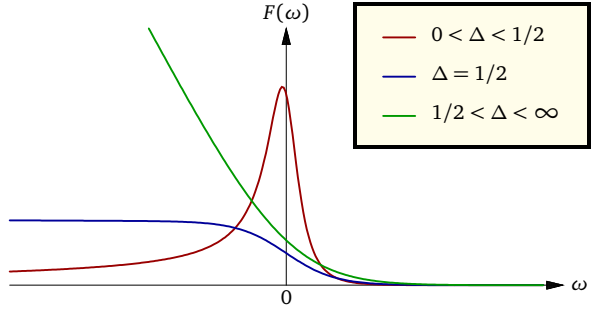}
  \caption{Typical detector's power spectra. The case
    $\omega=E_{f}-E_{i}>0$ corresponds to the absorption process while
    the case $\omega=E_{f}-E_{i}<0$ corresponds to the emission
    process. Note that the scaling dimension $\Delta$ is bounded below
    by $(d-2)/2$ in unitary conformal field theories.}
  \label{figure:1}
\end{figure}

Now, having the Wightman function for time-like separated operators,
one can easily find the Wightman function along the detector's
worldline \eqref{eq:3} by just replacing $-(x-y)^{2}$ with
$-(x(\tau)-x(\tau^{\prime}))^{2}=|\sinh(\pi
T(\tau-\tau^{\prime}))|^{2}/(\pi T)^{2}$. For $\tau\neq\tau^{\prime}$,
the result is
\begin{align}
  \langle0|\mathcal{O}_{\Delta}(x(\tau))\mathcal{O}_{\Delta}(x(\tau^{\prime}))|0\rangle
  &=\frac{\Gamma(2\Delta)A_{\Delta}}{2\pi}\mathrm{e}^{-i\,\mathrm{sgn}(\tau-\tau^{\prime})\pi\Delta}
    \left[\frac{\pi T}{\left|\sinh(\pi T(\tau-\tau^{\prime}))\right|}\right]^{2\Delta}\nonumber\\
  &=\frac{\Gamma(2\Delta)A_{\Delta}}{2\pi}\left[\frac{\pi T}{i\sinh(\pi T(\tau-\tau^{\prime}))}\right]^{2\Delta},\label{eq:16}
\end{align}
which, up to the overall normalization factor, exactly coincides with
the chiral half of the two-point Wightman function for
$(1+1)$-dimensional anyon field at finite temperature $T=a/(2\pi)$
\cite{Ilieva:1999bn,Liguori:1999tw,Ilieva:2000,Ilieva:2000cj,Ilieva:2000xt,Mintchev:2012pe},
where the proper time $\tau$ plays the role of the light-cone
coordinates $x^{\pm}=x^{0}\pm x^{1}$ of two-dimensional Minkowski
spacetime. In fact, eq.~\eqref{eq:16} has the desired properties for
anyon two-point Wightman functions at finite temperature:
\begin{itemize}
\item \textbf{Hermiticity.}
  \begin{align}
    \overline{\langle0|\mathcal{O}_{\Delta}(x(\tau))\mathcal{O}_{\Delta}(x(\tau^{\prime}))|0\rangle}
    =\langle0|\mathcal{O}_{\Delta}(x(\tau^{\prime}))\mathcal{O}_{\Delta}(x(\tau))|0\rangle
    \quad\text{for}\quad
    \forall\tau,\tau^{\prime}\in\mathbb{R}.\label{eq:17}
  \end{align}

\item \textbf{Anyon exchange relation.}
  \begin{align}
    \langle0|\mathcal{O}_{\Delta}(x(\tau))\mathcal{O}_{\Delta}(x(\tau^{\prime}))|0\rangle
    =\mathrm{e}^{-i\,\mathrm{sgn}(\tau-\tau^{\prime})2\pi\Delta}
    \langle0|\mathcal{O}_{\Delta}(x(\tau^{\prime}))\mathcal{O}_{\Delta}(x(\tau))|0\rangle
    \quad\text{for}\quad
    \tau\neq\tau^{\prime}.\label{eq:18}
  \end{align}

\item \textbf{KMS condition.}
  \begin{align}
    \langle0|\mathcal{O}_{\Delta}(x(\tau))\mathcal{O}_{\Delta}(x(\tau^{\prime}))|0\rangle
    =\langle0|\mathcal{O}_{\Delta}(x(\tau^{\prime}))\mathcal{O}_{\Delta}(x(\tau+\tfrac{i}{T}))|0\rangle
    \quad\text{for}\quad
    \forall\tau,\tau^{\prime}\in\mathbb{R}.\label{eq:19}
  \end{align}
\end{itemize}
Notice that the scaling dimension $\Delta$ plays the role of the
so-called statistical parameter $\kappa=2\Delta$, which interpolates
the Bose--Einstein and Fermi--Dirac statistics in a continuous
way. Indeed, the power spectrum $F(\omega)$ interpolates the
Bose--Einstein and Fermi--Dirac distributions. To see this, we need to
Fourier-transform the Wightman function \eqref{eq:16}, which can be
done as follows:
\begin{align}
  F(\omega)
  &=\int_{-\infty}^{\infty}\!\!\!d\tau\,
    \mathrm{e}^{-i\omega\tau}\langle0|\mathcal{O}_{\Delta}(x(\tau))\mathcal{O}_{\Delta}(x(0))|0\rangle\nonumber\\
  &=\frac{\Gamma(2\Delta)A_{\Delta}}{2\pi}\int_{-\infty}^{\infty}\!\!\!d\tau\,
    \mathrm{e}^{-i\omega\tau}\left[\frac{\pi T}{i\sinh(\pi T(\tau-i\epsilon))}\right]^{2\Delta}\nonumber\\
  &=\frac{\Gamma(2\Delta)A_{\Delta}}{2\pi}\int_{-\infty-\frac{i}{2T}}^{\infty-\frac{i}{2T}}\!\!\!d\tau\,
    \mathrm{e}^{-i\omega\tau}\left[\frac{\pi T}{i\sinh(\pi T\tau)}\right]^{2\Delta}\nonumber\\
  &=\frac{\Gamma(2\Delta)A_{\Delta}}{2\pi}(2\pi T)^{2\Delta-1}\mathrm{e}^{-\frac{\omega}{2T}}\int_{0}^{\infty}\!\!\!dz\,
    \frac{z^{\Delta-\frac{i\omega}{2\pi T}-1}}{(1+z)^{2\Delta}}\nonumber\\
  &=\frac{A_{\Delta}}{2\pi}(2\pi T)^{2\Delta-1}\mathrm{e}^{-\frac{\omega}{2T}}
    \left|\Gamma\left(\Delta+\frac{i\omega}{2\pi T}\right)\right|^{2},\label{eq:20}
\end{align}
where in the third equality we have pushed the integration contour
slightly downward by $i/(2T)$ in such a way that the integrand does
not blow up along the contour. The fourth equality follows from the
change of variable $z=\mathrm{e}^{2\pi T(\tau+i/(2T))}$ and the last
equality follows from the integral expression for the beta function:
\begin{align}
  B(p,q)
  =\int_{0}^{\infty}\!\!\!dz\,\frac{z^{p-1}}{(1+z)^{p+q}}
  =\frac{\Gamma(p)\Gamma(q)}{\Gamma(p+q)}
  \quad\text{for}\quad
  \mathrm{Re}\,p>0~~\&~~\mathrm{Re}\,q>0,\label{eq:21}
\end{align}
where $p=\Delta-i\omega/(2\pi T)$ and $q=\Delta+i\omega/(2\pi
T)$. Notice that $|\Gamma(x+iy)|^{2}=\Gamma(x+iy)\Gamma(x-iy)$ for
$x,y\in\mathbb{R}$. Note also that the power spectrum \eqref{eq:20}
satisfies the KMS condition in frequency space:
\begin{align}
  F(-\omega)=\mathrm{e}^{\omega/T}F(\omega).\label{eq:22}
\end{align}
In order to see the statistics interpolation, we first note that, just
as in eqs.~\eqref{eq:10a} and \eqref{eq:10b}, the power spectrum
\eqref{eq:20} can be recast into the following suggestive forms for
any $\Delta$:
\begin{align}
  F(\omega)
  =A_{\Delta}\frac{(2\pi T)^{2\Delta}}{\omega}
  \left|\frac{\Gamma\left(\Delta+\frac{i\omega}{2\pi T}\right)}{\Gamma\left(\frac{i\omega}{2\pi T}\right)}\right|^{2}
  \frac{1}{\mathrm{e}^{\omega/T}-1}
  =A_{\Delta}(2\pi T)^{2\Delta-1}
  \left|\frac{\Gamma\left(\Delta+\frac{i\omega}{2\pi T}\right)}{\Gamma\left(\frac{1}{2}+\frac{i\omega}{2\pi T}\right)}\right|^{2}
  \frac{1}{\mathrm{e}^{\omega/T}+1},\label{eq:23}
\end{align}
which follows from the Euler's reflection formulae
$|\Gamma(ix)|^{2}=\pi/(x\sinh(\pi x))$ and
$|\Gamma(1/2+ix)|^{2}=\pi/\cosh(\pi x)$ for $x\in\mathbb{R}$. Now it
is easy to see that, when $\Delta$ is either an integer or a half-odd
integer, the Gamma function factors reduce to polynomials of $\omega$.
In these cases the power spectrum results in the following:
\begin{align}
  F(\omega)=
  \begin{cases}
    \displaystyle
    \frac{A_{\Delta}\omega^{2\Delta-1}}{\mathrm{e}^{\omega/T}-1}\prod_{n=1}^{\Delta-1}\left[1+\left(\frac{2n\pi
          T}{\omega}\right)^{2}\right]
    &\text{for}\quad\Delta=1,2,\cdots,\\[2em]
    \displaystyle
    \frac{A_{\Delta}\omega^{2\Delta-1}}{\mathrm{e}^{\omega/T}+1}\prod_{n=\frac{1}{2}}^{\Delta-1}\left[1+\left(\frac{2n\pi
          T}{\omega}\right)^{2}\right]
    &\text{for}\quad\Delta=\frac{1}{2},\frac{3}{2},\cdots,
  \end{cases}\label{eq:24}
\end{align}
where for $\Delta=1$ and $1/2$ the products should be
disregarded. These give the explicit expressions of the power spectrum
\eqref{eq:9} for massless free scalar fields if the scaling dimension
is $\Delta=(d-2)/2$, $d\in\{3,4,5,\cdots\}$. Hence for generic
$\Delta$, the power spectrum \eqref{eq:20} interpolates the
Bose--Einstein and Fermi--Dirac distributions continuously. In fact,
it coincides with the so-called momentum distribution of
$(1+1)$-dimensional chiral anyons at finite temperature
\cite{Liguori:1999tw,Mintchev:2012pe}. Typical power spectra are
depicted in figure \ref{figure:1}.

To summarize, we have seen that, as a consequence of the branch point
singularity, the conformal two-point function generally obeys the
anyon(-like) exchange relation in the time-like domain. Its Fourier
transform along the worldline of uniformly accelerating detector then
coincides with the thermal distribution for $(1+1)$-dimensional anyons
at the Unruh temperature $T=a/(2\pi)$. It is worth mentioning here
that the detector's power spectrum enjoys yet another geometrical
interpretation. As briefly noted in
refs.~\cite{Takagi:1986kn,Ooguri:1985nv}, the Fourier transform
$F(\omega)$ is related to the nontrivial density of states on the
Rindler wedge. In fact, for the case of canonical dimension
$\Delta=(d-2)/2$, the power spectrum is proportional to the Plancherel
measure for a scalar field on the $(d-1)$-dimensional hyperbolic space
$\mathbb{H}^{d-1}$ (see, e.g., eq.~(2.36) in
ref.~\cite{Camporesi:1994ga}):\footnote{Notice that the
  $d$-dimensional Rindler wedge is conformal to
  $\mathbb{H}^{1}\times\mathbb{H}^{d-1}$; see, e.g.,
  ref.~\cite{Ohya:2016gto}.}
\begin{align}
  \mu_{0}(\lambda)\propto\left|\frac{\Gamma((d-2)/2+i\lambda)}{\Gamma(i\lambda)}\right|^{2},\label{eq:25}
\end{align}
where $\lambda$ should be read as $\omega/(2\pi T)$. Obviously,
eq.~\eqref{eq:25} can be analytically continued to arbitrary
$\Delta=(d-2)/2$. Hence the detector's power spectrum \eqref{eq:23}
allows two distinct interpretations: one is the Bose--Einstein
distribution multiplied by the Plancherel measure, and the other is
the anyon distribution. The Takagi's statistics inversion thus comes
from the peculiar property of this Plancherel measure. It is quite
interesting that the Unruh effect gives a geometrical interpretation
for the anyon distribution and may well provide a surprising
connection between statistics and geometry.

\subsection*{Acknowledgments}
\label{acknowledgements}
The author would like to thank Mihail Mintchev for discussions and
comments. He is also grateful to INFN Pisa for hospitality during his
visit at the University of Pisa. This work was supported in part by
JSPS Grant-in-Aid for Research Activity Startup \#15H06641.

\bibliographystyle{utphys}%
\bibliography{bibliography}%
\end{document}